\documentclass[fleqn,10pt]{wlscirep}
\usepackage[utf8]{inputenc}
\usepackage[T1]{fontenc}
\usepackage{afterpage}

\usepackage{booktabs}
\usepackage{multirow}
\usepackage{longtable}
\usepackage{xr-hyper}
\usepackage[charter,cal]{mathdesign}

\makeatletter
\newcommand*{\addFileDependency}[1]{
  \typeout{(#1)}
  \@addtofilelist{#1}
  \IfFileExists{#1}{}{\typeout{No file #1.}}
}
\makeatother

\newcommand*{\myexternaldocument}[1]{%
    \externaldocument{#1}%
    \addFileDependency{#1.tex}%
    \addFileDependency{#1.aux}%
}
\myexternaldocument{si}

\title{
The role of central places in exposure segregation
}
\author[a,*]{Andrew Renninger}
\author[a]{Mateo Neira}
\author[a]{Elsa Arcaute}

\affil[a]{Centre for Advanced Spatial Analysis, University College London, London, UK}

\affil[*]{Corresponding author: Andrew Renninger (E-mail: andrew.renninger.12@ucl.ac.uk)}

\begin{abstract}
Here we show that ``exposure segregation"—the degree to which individuals of one group are exposed to individuals of another in day-to-day mobility—is dependent on the structure of cities, and the importance of downtowns in particular. Recent work uses aggregated data to claim that the location of amenities can inhibit or facilitate interactions between groups: if a city is residentially segregated, as many American cities are, then amenities \emph{between} segregated communities should encourage them to mix. We show that the relationship between ``bridging" amenities and socio-economic mixing breaks down when we examine the amenities themselves, rather than the urban aggregates. For example, restaurants with locations that suggest low expected mixing do not, much of the time, have low mixing: there is only a weak correlation between bridging and mixing at the level of the restaurant, despite a strong correlation at the level of the supermarket. This is because downtowns—and the bundle of amenities that define them–tend not to be situated in bridge areas but play an important role in drawing diverse groups together. 
\end{abstract}

\begin{document}

\flushbottom
\maketitle

\section*{Introduction}
New research finds that the distribution of amenities in a city relates to socio-economic mixing: cities with ``hubs" situated between communities of different socio-economic strata tend to have more exposure across those strata; cities with hubs located comfortably within homogeneous zones have less \cite{nilforoshan2023human}. Using data from the same provider, the following analysis shows that although hubs are associated with mixing across metropolitan areas, when we sharpen the spatial resolution of the analysis, the association attenuates considerably, raising concerns about causality. Further, this relationship ignores the importance of hierarchy in urban structure first articulated in Central Place Theory, an idea which began in economic geography and has found support in recent studies of mobility. In brief, this theory posits a hierarchy of locations wherein some goods or services are provided frequently through dispersed vendors and others—central goods—are accessed infrequently through centralized agglomerations of amenities \cite{christaller1966central}. We can improve our understanding of mixing—and with it our policy recommendations—by first incorporating and then validating ideas on central places and central goods.

Without explicit consideration for the role of central places and central goods in driving mobility, a growing body of literature indicates that the location and co-location of amenities in space is the product of related forces. Consistent with the idea that certain goods are traded centrally and others are served in a dispersed manner, with consequences for how often transactions take place, data on mobility show that visits to an area are a function of frequency and distance: some places are visited often by those nearby, and other places are visited sparingly by people from far afield \cite{schlapfer2021universal}. Other work has revealed distinct hierarchies in travel, with trips bounded by local or regional limits \cite{alessandretti2020scales}, and constraints on the number of places any given person can visit \cite{alessandretti2018evidence}. Together, much of the work on human mobility in the urban context suggests that individuals must allocate finite resources—time, money and even cognitive bandwidth \cite{bongiorno2021vector, balaguer2016neural}—to satisfy needs. ``Trip chaining"—whereby people visit multiple services on the same journey—is a common strategy for coping with these limitations \cite{miyauchi2021economics}. Taken together, this research suggests that urban structure constrains human behavior, and vice versa—with amenities agglomerating to meet certain needs. Here we examine the implications of this interplay between structure and mobility on socio-economic mixing. 

\section*{Results}
We use a measure of mixing borrowed from earlier work \cite{moro2021mobility} that captures the degree to which the distribution of visitors to a place deviates from an ideal wherein visitors from each quintile of the income distribution visit that place in equal proportions. \footnote{This approach captures the distribution of visitors, rather than simple averaging \cite{nilforoshan2023human}, which has documented limitations when people belong to multiple areal units—as we would expect in cities where people can live in a dormitory community and work in a business district. For a detailed explanation of why this is our preferred measure, see the supplemental materials.} These quintiles are calibrated using the metropolitan area, rather than the nation as a whole, to compensate for the concentration of wealth in some cities. The result is a value from 0 to 1 where 0 represents perfect mixing and 1 represents perfect segregation. With that as our proxy for mixing computed for $\sim5$ million points of interest nationally, we construct our bridging index—which represents mixing under counterfactual mobility scenario—using a modified version of an established model \cite{simini2012universal} that assigns trips between origins and destinations considering the distance between them and the intervening opportunities; in this model, the probability that you visit a restaurant is a function of how far away that restaurant is and how many other restaurants are closer. We can think of this as an ``expected'' mixing to which we can compare observed data. We resample the observed data accordinging to these probabilities for many business types. The above mixing measure, computed on these assigned trips, constitute our bridging index; this counterfactual mixing indicates the potential for amenities to attract diverse patrons. 

\begin{figure*}[bt!]
\centering
\makebox[\textwidth][c]{\includegraphics[width=1\textwidth]{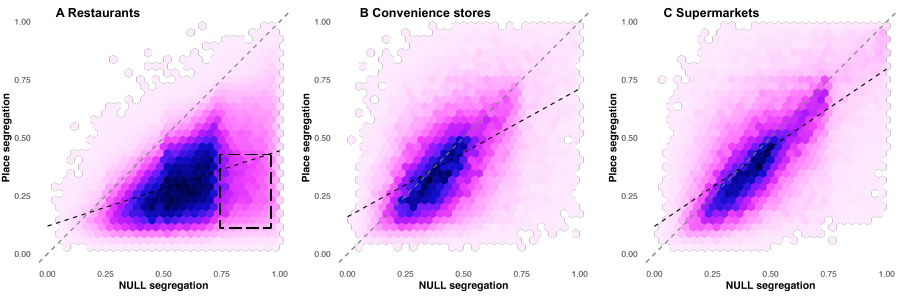}}
\caption{Plots showing the correlation between bridging and mixing at the level of amenity for different venue classes, with weaker correlation for \textbf{A} restaurants and stronger correlation for \textbf{C} supermarkets, and \textbf{B} convenience stores in the middle. A large portion of restaurants are high expected segregation and low observed segregation.}
\label{correlations}
\end{figure*}

Replicating earlier bridging-mixing findings \cite{nilforoshan2023human}, we find that our measures are well calibrated: bridging at the level of the city is correlated with mixing at the city as a whole ($\rho=0.63$) and each of our measures for bridging and mixing are correlated with those of the earlier study ($\rho=0.55$ and $\rho=0.57$, respectively). Yet this relationship varies when we disaggregate to level of the point of interest: some amenities encourage mixing and others do not. Looking at restaurants, the Pearson correlation coefficient for mixing and bridging falls to $0.40$ when we consider restaurants in Fig. ~\ref{correlations}\textbf{A}. At supermarkets, however, which are fewer in number and perhaps less separable according to taste and preference, the correlation coefficient between bridging and mixing is $0.70$ (see Fig. ~\ref{correlations}\textbf{C}). Convenience stores are between these 2 extremes, with a correlation of $0.53$ (see Fig. ~\ref{correlations}\textbf{B}). There is a notable surplus of restaurants in particular off the diagonal, where we would expect less mixing than occurs.

This presents a puzzle that calls into question the causal effect of bridging on mixing in American cities: cities with bridging amenities experience less exposure segregation than those without, but many amenities with low bridging also have high mixing. Our data focuses on venues, so it is plausible that much of the mixing in cities occurs on the street in front of restaurants and grocers as the patrons themselves sort into amenities along socio-economic characteristics. This sorting would align with work showing sorting among nearby venues across cities \cite{moro2021mobility}, and would indicate that exposure is superficial—shared paths rather than amenities. It is also plausible that unobserved variables produce both bridging and mixing, and that the two are not causally linked. We check our hypothesis that macro-mixing occurs alongside micro-sorting by aggregating venues into clusters and comparing cluster-level mixing to amenity-level mixing; we also check for systematic biases in the relationship between bridging and mixing. Fig.~\ref{hubs} presents these results.

\afterpage{%
\clearpage
\begin{figure*}[h!]
\centering
\makebox[\textwidth][c]{\includegraphics[width=1\textwidth]{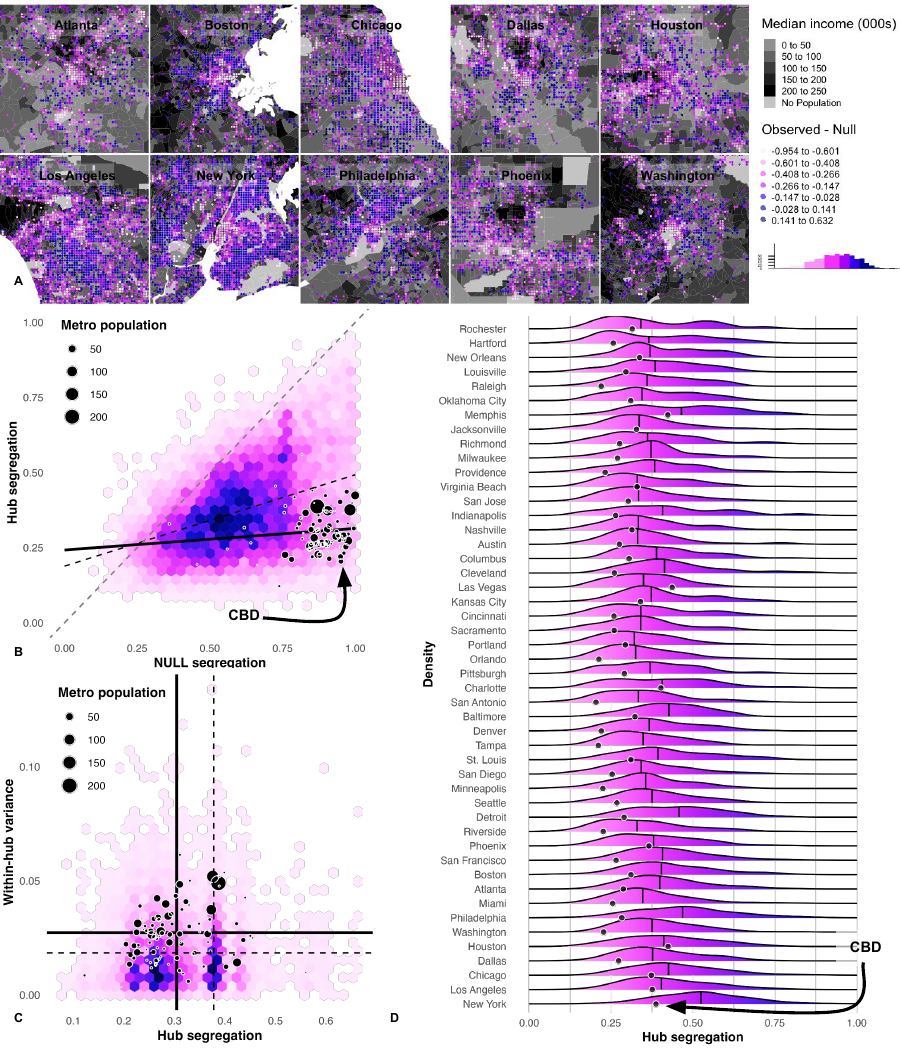}}
\caption{\textbf{A} Maps illustrating the difference between bridging potential and place segregation, showing distinct zones of mixing—usually downtown—and segregation. \textbf{B} Plots showing the correlation between bridging and mixing at the level of the hub, with downtown clusters extracted. \textbf{C} Heightened mixing in these clusters does not come with greater variance, generally, which would suggest that people are not sorting within clusters. \textbf{D} Distributions for all hubs across the top 50 cities by population, with the downtown cluster extracted, shows that these clusters tend to be at the integrated tail of the distribution.}
\label{hubs}
\end{figure*}
\clearpage
}

When we subtract expected segregation from observed segregation, we see that the lowest values—segregated well below expectation, given location—are concentrated downtown (see: Fig.~\ref{hubs}\textbf{A}). Further, the relationship between bridging and mixing breaks down when considering these business districts: these hubs typically have high predicted segregation—because they are situated far away from much of the broader suburban and exurban population—and low observed segregation. In Fig.~\ref{hubs}\textbf{B}, we compute hub mixing across all clusters of amenities and then extract downtowns, which we define as the largest cluster in each city, to demonstrate. This also explains the weak relationship between mixing and bridging for restaurants: many restaurants are co-located in larger business clusters; these restaurants tend to be the low-bridging, high-mixing venues we see above. Considering the 2 large clusters in New York City and Chicago in particular, this relationship is partially driven by sorting within the cluster at the level of the venue. In Fig.~\ref{hubs}\textbf{C}, We use the variance in median income across establishments to illustrate: CBDs tend to have lower segregation on aggregate but higher variance between venues, indicating sorting within them. Many of the most segregated CBDs are in cities with even more segregation, and are thus still doing comparably better: as shown in Fig.~\ref{hubs}\textbf{D}, downtowns are typically far to the integrated side of their parent city's distribution.

\begin{figure*}[ht!]
\centering
\makebox[\textwidth][c]{\includegraphics[width=0.5\textwidth]{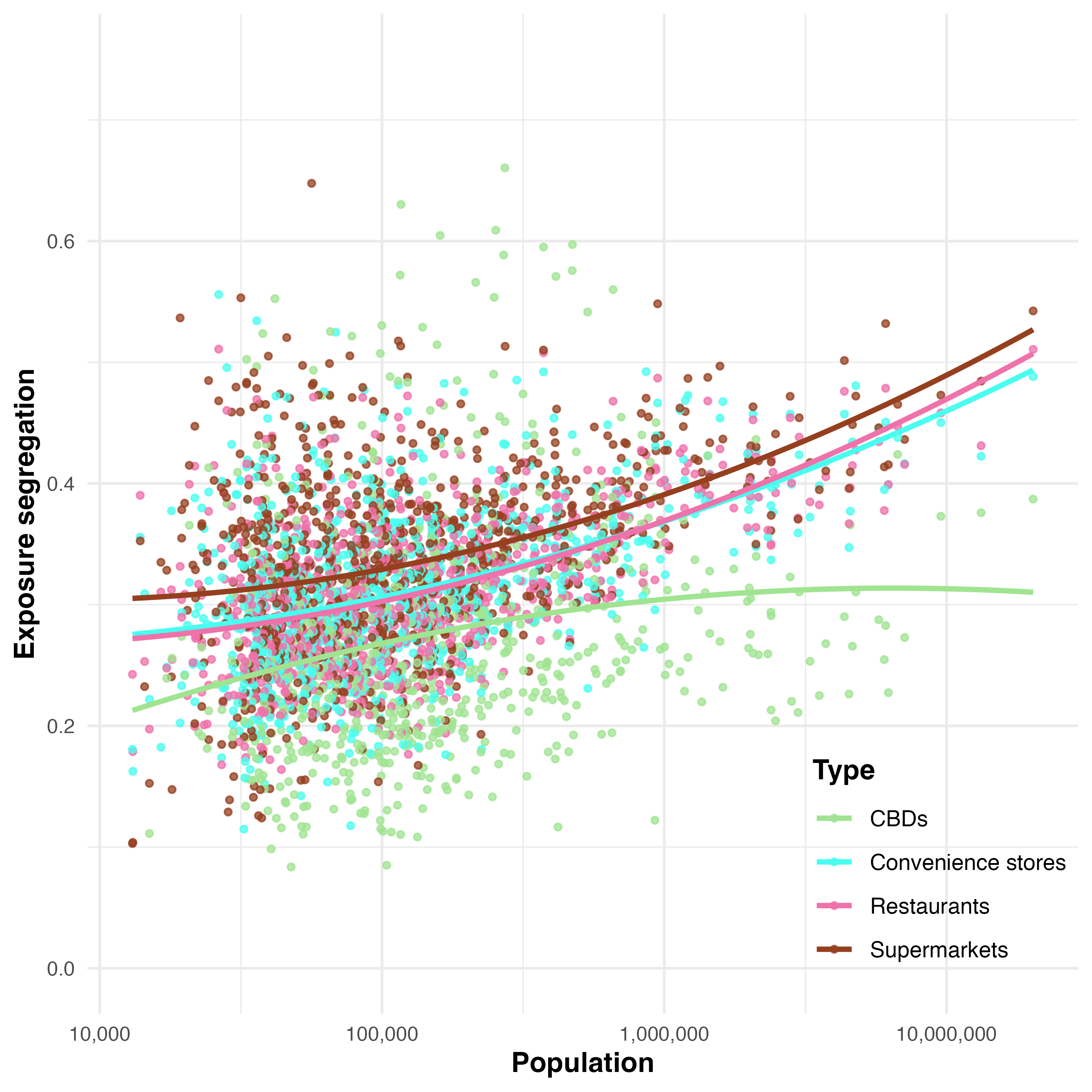}}
\caption{On an aggregate level as well as for certain kinds of amenities, exposure segregation scales with city size, but this relationship breaks down when we isolate central business districts.}
\label{scaling}
\end{figure*}

The consequence of this is that the scaling relationship found in prior work \cite{nilforoshan2023human}, which says that larger cities have greater exposure segregation, does not hold when we isolate central business districts. Fig. \ref{scaling} shows that scaling holds for each type of business we explore here, but not for CBDs. This suggests that much of changes to exposure segregation across cities of different sizes is due to changes at the periphery. 

\section*{Discussion}
In this brief communication we find that the intrinsic mixing potential of an amenity is dependent on its type and its position in the urban hierarchy. Prior work has shown that cities with amenities distributed to ``bridge" demographic divides in the population experience more mixing. Yet these results ignore the importance of hierarchy in determining both human mobility and urban structure. Hierarchy is evident in the importance of downtowns, which tend to have low ``bridging" but high mixing—for the simple reason that they tend be located in the center of concentric zones sorted by race, class and age, rather than between them. Downtown offices draw commuters past intervening amenities, generating mixing in a unified central location. This has the strongest effect on restaurants, which are \emph{dispersed} rather than \emph{central} amenities—those for which patrons would prefer not to travel long distances, but end up doing so when that visit is attached to a commute to the office—a \emph{central} place. Supermarkets are not typically part of downtown amenity clusters and for them bridging does indeed predict mixing. Comparing supermarkets stores with convenience stores, we see that the comparably more central supermarkets show a tighter bridging-mixing link than dispersed convenience stores, which suggests that location is more important to the mixing profile of anchor businesses that are attractions in and of themselves than businesses that attach to larger clusters. These results will have particular relevance as many jobs switch to hybrid or remote work, curbing the importance of urban downtowns in the urban hierarchy \cite{ramani2021donut} and placing greater demand on suburban attractions. 

\section*{Methods}
We construct our measure for socio-economic mixing with data from SafeGraph \cite{safegraph2019visitation}, a location services provider, which comes in the form of origin-destination flows from neighborhoods (Census block groups) to points of interest. SafeGraph gathers location data from smart phones by aggregating GPS logs from applications that have obtained user consent to collect such data; the sample constitutes $\sim10\%$ of the population. They assign visits to points of interest by clustering GPS pings and joining these clusters to adjacent building polygons, using relative distances and time-of-day to manage conflicts \cite{safegraph2019visitation}. Previous work has shown that SafeGraph data are demographically calibrated \cite{squire2019bias}, and this data underlies the study on mixing and bridging referenced here \cite{nilforoshan2023human}. We use this rectangular matrix consisting of 220,000 origin home block groups (a Census aggregation with a population $\sim1000$) and 7 million destination points of interest like restaurants, grocers, and other businesses. We infer the socio-economic strata of the visitors from each origin with Census estimates from the American Community Survey \cite{USCensusBureau_ACS} of median income in that area.

\paragraph{Measuring mixing.} We measure mixing by looking at a measure of balance proposed in \cite{moro2021mobility} that asks, in what proportions do visitors from different socio-economic strata visit an amenity? If a restaurant attracts visitors from just one income bracket, we consider that amenity to be segregated; if it attracts visitors from different brackets in equal proportion, then we consider it to be mixed. Following the convention of earlier work, we consider the segregation $S$ of an amenity $\alpha$ to be a distance from an ideal scenario where people from all socio-economic classes visit in equal proportions. This is defined as follows

\begin{equation} \label{eq:1}
S_{\alpha} = \frac{5}{8} \sum_{q} \left| \nu_{q\alpha} - \frac{1}{5} \right|,
\end{equation}

\noindent
where $q$ represents an income quintile and $\nu$ represents the portion of visitors from that quintile. We scale by $\frac{5}{8}$ so that each value spans 0 to 1, with 0 being perfect integration (equal proportions from all classes) and 1 being perfect segregation (visitors from a single class). Each quintile is calibrated to the metropolitan area, rather than the nation as a whole.

\paragraph{Computing bridging.} We are interested in generating an expected segregation estimate to compare to the observed. To do this, we model flows to amenities using a modified radiation model \cite{simini2012universal}, which generates flows from origins to destinations according to intervening opportunities, and then compute the same segregation measure on the simulated data. A radiation model typically uses areal units like counties or tracts as origins and destinations; here we have origin neighborhoods represented by census block groups and destination amenities, stratified by business type (defined by 6-digit NAICS code) so that we are not apportioning visits between, for example, restaurants and grocers. Because of this we recast this model as one that asks, how far is origin $i$ from destination $j$ and how many comparable (same NAICS) destinations are between $i$ and $j$? We define probability of interaction between origin $i$ and destination $j$ as

\begin{equation} \label{eq:2}
P_{ij} = \frac{e^{-\beta d_{ij}}}{(1 + r_{ij}) \sum_{k} \frac{e^{-\beta d_{ik}}}{r_{ik}}},
\end{equation}

\noindent
where $P_{ij}$ is the probability of interaction between origin $i$ and destination $j$, $\beta$ is the decay parameter (we use 0.00005, which best fits the data, but our results hold for 0.00001 and 0.00010), $d_{ij}$ is the distance between origin $i$ and destination $j$, $r_{ij}$ is the ranked distance of destination $j$ from origin $i$. The sum in the denominator is over all destinations $k$ for origin $i$. We then resample the total observed trips emanating from each origin according to that probability distribution. Null segregation is simply computed for these results with eq. \ref{eq:1}. 


\bibliography{exposure}

\section*{Acknowledgements}
The authors would like to thank the members of our research group who contributed comments throughout the process.

\section*{Author contributions statement}
\textbf{A.R.} and \textbf{M.N.} Conceptualization, methodology, investigation, writing, reviewing, editing; \textbf{E.A.} Supervision, reviewing and editing.  

\subsection*{Data and code availability}
Owing to the detailed nature of our analysis, we will provide expected and observed segregation data with deidentified points of interest upon request.

\section*{Competing interests}
The authors declare no conflict of interest.

\end{document}


\flushbottom
\maketitle

Recent research uses the correlation between a person's socio-economic status (SES) and the mean SES of their exposure network as a measure of segregation to  challenge the "cosmopolitan mixing hypothesis" \cite{nilforoshan2023human}. The measure is formalized as follows:

\begin{equation}\label{eq:exposure_segregation}
    \text{Exposure Segregation} = \text{Corr}(SES, \overline{\text{SES}}_{\text{exposures}} ) = \frac{\text{cov}(SES, \overline{ \text{SES} }_{\text{exposures}})}{\sigma_{SES} \times \sigma_{ \overline{ \text{SES} }_{\text{exposures}}}}
\end{equation}
here $\text{cov}(SES, \overline{ \text{SES} }_{\text{exposures}})$ represents the covariance between an individual's SES and the SES of people they encounter, while $\sigma_{SES} \times \sigma_{ \overline{ \text{SES} }_{\text{exposures}}}$ are the standard deviations of these SES values.

The claim is that this \textit{exposure segregation} measure is a generalization of a widely used measure of socioeconomic segregation — the neighbourhood sorting index \cite{jargowsky1996take}. 

Here, we argue that although the neighbourhood sorting index is equivalent to calculating the Pearson correlation coefficient ($\rho$) between a person's SES and the SES of the spatial unit they reside in, this equivalency only holds because each individual belongs to exactly one spatial unit, and their SES contributes to the mean of that unit alone. This one-to-one relationship breaks down in the case of the exposure network, where an individual can contribute to the mean of multiple groups simultaneously. Because of this, the Pearson correlation between individual's SES and the mean SES's of their exposure network throws away important information about the within group variance, which is crucial for quantifying segregation.

\section*{Equivalence of the correlation ration ($\eta^2$), the neighbourhood sorting index (NSI) and Pearson correlation coefficient ($\rho$)}

The problem of measuring segregation by a continuous variable, like income, has been widely studied. The \textit{correlation ration} ($\eta^2$) \cite{duncan1955methodological}, has been often used to measure of segregation that is well-defined for continuous variables. This measure provides a principled way to quantify segregation that is independent of mean and variance of individual income. Here we first show the equivalence of $\eta^2$ to $NSI$ \cite{jargowsky1996take}, then we show that the $NSI$ is equivalent to calculating $\rho$ between individual incomes and the mean income of the spatial unit they reside in. Lastly, we show that this equivalence does not hold when individuals can be part of multiple groups simultaneously.

If we have a total $N$ individuals distributed in $J$ spatial units, where $N_j$ is the total number of individuals in the $j$-th spatial unit, $x_i$ represents the wealth of the $i$-th individual, $\bar{x}$ the mean wealth of the entire population and $\bar{x_j}$ the mean wealth of individuals in the $j$-th spatial unit, then:

\begin{equation}\label{eq:cr}
\eta^2 = 1 - \frac{ \sqrt{ \sigma_W^2} }{ \sqrt{ \sigma_N^2} }
\end{equation}
where $\sigma_W^2$ (eq. \ref{eq:within_variance}) is the within-spatial-unit variance of wealth and $\sigma_N^2$ (eq. \ref{eq:total_variance}) is the total variance of wealth of the population. 

\begin{equation}\label{eq:within_variance}
\sigma_W^2 = \frac{1}{N} \sum_{j=1}^{J} \sum_{i=1}^{N_j} (X_{ij} - \bar{X_j})^2 
\end{equation}

\begin{equation}\label{eq:total_variance}
\sigma_N^2 =  \frac{1}{N} \sum_{i=1}^{N} (X_{i} - \bar{X})^2 
\end{equation}
if individuals where randomly distributed among spatial units the within-spatial-unit variance would approximate the total variance and would result in $\eta^2$ close to zero, indicating no spatial segregation. If, on the other hand, spatial units where internally homogeneous in the wealth of individuals but the mean wealth of each spatial unit was different, then the within-spatial-unit variance would be small compared to the total variance and the resultant $\eta^2$ would be close to one, indicating high levels of segregation. 

The neighbourhood sorting index \cite{jargowsky1996take} is a mathematically equivalent formulation of $\eta^2$ (eq. \ref{eq:eta} ) that uses the between-spatial-unit variance $\sigma_B^2$ (eq. \ref{eq:between_variance}) instead of $\sigma_W^2$. 

\begin{equation}\label{eq:eta}
NSI = \frac{ \sqrt{ \sigma_B^2}}{ \sqrt{\sigma_N^2} }
\end{equation}
where: 
\begin{equation}\label{eq:between_variance}
\sigma_B^2 =  \frac{1}{N} \sum_{j=1}^{J} N_j (X_{j} - \bar{X})^2 
\end{equation}
The two formulations are equivalent because the total variance $\sigma_N^2$ is the sum of $\sigma_B^2$ and $\sigma_W^2$. The total variance can be decomposed in this way because spatial units are independent of each other, resulting in non-overlapping within-spatial-unit variance. 

The NSI can be calculated by the Pearson correlation coefficient of each individual's income and the mean income of their spatial unit $ \rho_{x_{i,j}, \bar{x_j}} $. In this context,  $ \rho_{x_{i,j}, \bar{x_j}} $ represents the total proportion of variance in individual income $x_{i,j}$ that is associated with the mean wealth of the spatial unit $\bar{x_j}$, quantifying the proportion of the total variance explained by the spatial unit categorization, thus equivalent to $NSI$.

In the case of residential segregation, as shown above, $ \rho_{x_{i,j}, \bar{x_j}} $ is able to capture the relationship between within group and total group variance, providing a principled approach to quantifying segregation with continuous variables. However, this only holds because of the unique individual-to-spatial unit relationship that allows the simple decomposition of the total variance into within and between spatial unit variance. When we move to the exposure network case, where an individual can belong to multiple groups simultaneously, resulting in group overlap, this simple decomposition is no longer possible and the resultant $ \rho_{x_{i,j}, \bar{x_j}} $ does not cleanly capture the relationship between within-group and total variance that is necessary to fully quantify segregation.


\bibliographystyle{naturemag}
\bibliography{exposure}